\documentclass[twocolumn,showpacs,preprintnumbers,amsmath,amssymb,prl]{revtex4}

\usepackage{graphicx}
\usepackage{dcolumn}
\usepackage{bm}
\newcommand{\be}{\begin{equation}}
\newcommand{\ee}{\end{equation}}
\newcommand{\bea}{\begin{eqnarray}}
\newcommand{\eea}{\end{eqnarray}}
\newcommand{\df}{{\rm d}}

\begin{document}

\preprint{}

\title{Embedding a chaotic signature in a periodic train: can periodic signals be chaotic?}

\author{Antonio Mecozzi}%
\email{amecozzi@ing.univaq.it}
\author{Cristian Antonelli}
\affiliation{%
Dipartimento di Ingegneria Elettrica e dell'Informazione and CNISM, Universit\`a dell'Aquila, Poggio di Roio, I-67040 L'Aquila, Italy
}%

\date{\today}

\begin{abstract}

We show how a chaotic system can be locked to emit a periodic waveform belonging to its chaotic attractor. We numerically demonstrate our idea in a system composed of a semiconductor laser driven to chaos by optical feedback from a short external cavity. The clue is the injection of an appropriate periodic signal that modulates the phase and amplitude of the intra-cavity radiation, a chaotic analogy of conventional mode-locking.  The result is a time process that manifests a chaotic signature embedded in a long-scale periodic train.

\end{abstract}



\maketitle

Chaos is an important behavior of dynamical systems that arises in dire distinction to standard Newton-like behavior in a great variety of physical arenas \cite{Ruelle}. Chaos has been observed, for instance, in electrical circuit, fluid dynamics and laser. Chaotic behavior is a phase of the dynamics, and it is  an accepted notion that a system that is chaotic is not periodic, and viceversa.  Periodicity and chaos are, however, closely related.  For example, a stable system can be routed to chaos passing through a limit cycle, which represents a periodic trajectory in phase space. The regularity of a limit cycle is however a distinct signature that makes it easily distinguishable from a chaotic trajectory of the system.

In this paper we show that it is possible to stabilize a periodic trajectory that preserves the qualitative features of a chaotic trajectory by injection of a periodic waveform. Our scheme is different from previous schemes of stabilization of a periodic trajectory in dynamical systems \cite{Ohtsubo_PRE, Watanabe,Gauthier} in a way that in these previous examples the period of the stabilized trajectory is typically comparable to the time scales of the system dynamics, making the stabilized trajectory similar to a limit cycle more than to a chaotic trajectory. Here, on the other hand, the period of the stabilized trajectory is thousands of times longer than the typical time constants of the system dynamics and can be in principle arbitrarily long. This makes the stabilized periodic trajectory almost indistinguishable from a truly chaotic trajectory of the system, and rigorously indistinguishable if the period of the control signal tends to infinity. We conjecture that the ensemble of periodic trajectories that may be stabilized has a continuous transition to the entire ensemble of chaotic trajectories as the period of the control waveform tends to infinity. Our specific example refers to the case of a semiconductor laser driven to chaos by optical feedback from a short optical cavity  \cite{Ohtsubo}.

The dynamics of a semiconductor laser is described by the electric field of complex envelope $E(t)$, normalized such that $|E(t)|^2$ is the intra-cavity photon number, and by the carrier number $N(t)$. These quantities are governed by a modified set of Lang-Kobayashi (L-K) equations  \cite{L-K}, which we generalize including an extra term accounting for the injection of an external driving field,
\bea \frac{\df E}{\df t} &=& (1 + i \alpha) \left[G(t) - \frac{1}{ \tau_p} \right] \frac{E(t)}{2} \nonumber \\ && + \sqrt{\eta} \,  E(t - \Delta t) e^{i \omega_0 \Delta t} /\tau_{\textrm{in}} + E_{\mathrm{ext}}(t)/\tau_{\textrm{in}}, \label{10} \\ \frac{\df N}{\df t} &=& \frac{I}{e} - \frac{N(t)}{\tau_c} - G(t) |E(t)|^2, \label{20} \eea
with
\be G(t) = g \frac{N(t) - N_0}{1 + s |E(t)|^2}. \ee
Here, $N_0$ is the carrier number at transparency, $G(t)$ is the saturated gain, $g$ is the differential gain and $s$ is the gain suppression coefficient, $\alpha$ is the line-width enhancement factor, $\tau_p$ is the photon lifetime, $\tau_c$ is the carrier lifetime, $\tau_{\textrm{in}}$ is the round-trip time of the laser, $\eta$ is the fraction of the optical power fed-back to the laser cavity to drive the laser to chaos, $\Delta t$ is the external cavity delay, $\omega_0$ is the optical carrier angular frequency, $I$ is the pump current and $e$ is the electron charge. The term $E_{\mathrm{ext}}(t)$ is the field amplitude of the injected external driving field after passing through the laser facet. Equations (\ref{10}) and (\ref{20}) have been extensively used to study the laser dynamics and, when compared with experimental results, they have passed the accuracy test in the totality of cases \cite{Ohtsubo,Parameters}.

\begin{figure}
\includegraphics[width=8cm]{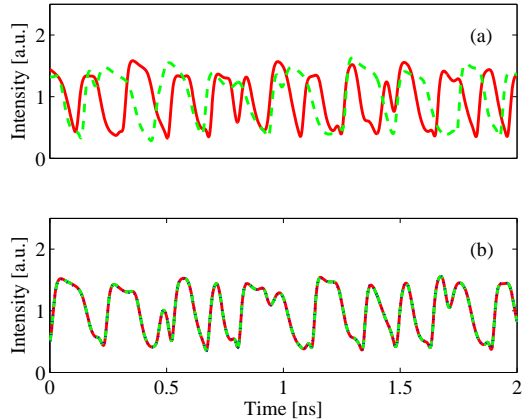}
\caption{Intensity $|E(t)|^2$ (dashed green curve) and $|E(t-T)|^2$ (solid red curve), normalized to unit average value, of the electric field at time $t$ and time $t-T$. Fig. \ref{Fig1}a refers to the free-running case, Fig. \ref{Fig1}b to the locked case, where the two curves overlap on the scale of the plot. In Fig. \ref{Fig1}b, the blue dotted line represents the normalized intensity of the driving field at both time $t$ and $t-T$.}
\label{Fig1}
\end{figure}

We use first for the forcing field $E_{\mathrm{ext}}(t)$ a replica of a portion of length $T$ of the chaotic intra-cavity field of the free-running laser, attenuated by the factor $\sqrt{\eta'}$ ($\eta'$ is the corresponding power fraction), periodically repeated with period $T$. We find that, after an initial transient of the order of a few $T$, the intra-cavity field locks into a periodic trajectory of period $T$. Results show stringent similarities with mode-locking. The periodic coherent injection of a delayed waveform of the same laser acts like the coupling with an ideal external cavity of mode spacing $1/T$, which is a well known technique to achieve mode-locking with the period of the external cavity round-trip time $T$ \cite{Ippen}. In this regime, the onset of a periodic evolution may be seen as a form of chaos self-synchronization \cite{Pecora}, or as a form of chaos control \cite{Ott}. In Fig. \ref{Fig1}a, we show by a solid red line the intensity of the field of the free-running laser, i.e. with $E_{\mathrm{ext}}(t)=0$, at time $t-T$, and with a dashed green line the same quantity at time $t$, normalized to unit average intensity. In Fig. \ref{Fig1}b, we report with a solid red line the intensity of the field at time $t-T$, with a dashed green line the same quantity at time $t$, and with a dotted blue line the intensity of the injected driving field $E_{\mathrm{ext}}(t)$, also normalized to unit average intensity. It is apparent that the laser intensity locks to the drive and into a periodic trajectory. We have verified that the full intra-cavity electric field, including the phase, and the carrier number are truly periodic, and that the periodic trajectory (once the laser parameters and the driving field are fixed) is a global attractor of the system, i.e. the system responds to a perturbation of electric field and carrier density returning to the same periodic trajectory.

\begin{figure}[t!]
\includegraphics[width=8cm]{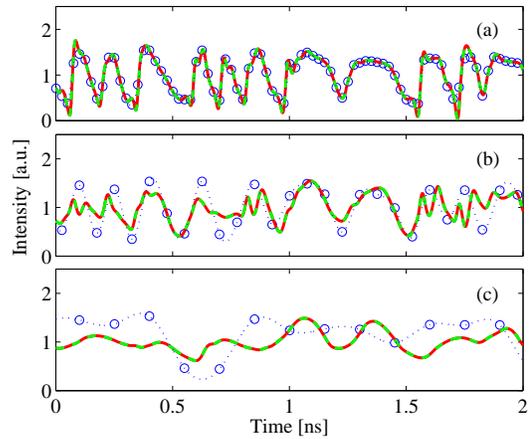}
\caption{Intensity $|E(t)|^2$ (dashed green curve) and $|E(t-T)|^2$ (solid red curve), normalized to unit average value, of the electric field at time $t$ and time $t-T$, together with the normalized intensity of the driving field (dotted blue curve), for a) $T_s = 25$ ps, (b) $T_s = 75$ ps, and c) $T_s = 150$ ps sampling time. Circles denote the sampling points.}
\label{Fig2}
\end{figure}

The values of the parameters of the numerical example are \cite{Parameters} $N_0 = 1.5\times10^8$, $g = 6\times10^{-6} \, \mathrm{ns}^{-1}$, $s = 5\times10^{-7}$, $\alpha = 4$, $\tau_p = 5$ ps, $\tau_c = 0.5$ ns, $\tau_{\textrm{in}} = 8$ ps, $\eta = 0.025$, $\Delta t = 3$ ns, $\omega_0 = 2 \pi c/\lambda$ with $c$ the speed of light and $\lambda = 750$ nm, $I = 1.5 I_{\mathrm{th}}$, with $I_{\mathrm{th}} = 60$ mA. We choose $\eta' = 0.16$, corresponding to a $16$ \% power coupling coefficient of the driving field. The period is $T=100$ ns.

In practice, the scheme described above may be realized in a master-slave configuration, by recording the complex chaotic waveform emitted by the free-running laser in a time interval $T$, and injecting in the same laser the output of a continuous-wave laser modulated in intensity and phase to periodically reproduce the recorded waveform. In real word, however, phase and amplitude modulation may only be applied within a finite bandwidth. For this reason, we decided to investigate the effect on the locking quality of the modulation bandwidth. For this purpose, we applied to the slave a repeated copy of the free-running chaotic intra-cavity field sampled with a variable sampling time $T_s$. The sampled points were connected by an interpolation of the sampled intensity and phase using a MatLab interpolation routine based on spline (i.e. cubic) functions. The results are shown in Fig. \ref{Fig2}, where we again compare the intensity of the field at time $t-T$ with the intensity at time $t$ and with the intensity, normalized as in Fig. \ref{Fig1}, of the forcing field $E_{\mathrm{ext}}(t)$, for $T_s = 25$ ps, $T_s = 75$ ps, and $T_s = 150$ ps sampling time.

\begin{figure}[t!]
\includegraphics[width=8cm]{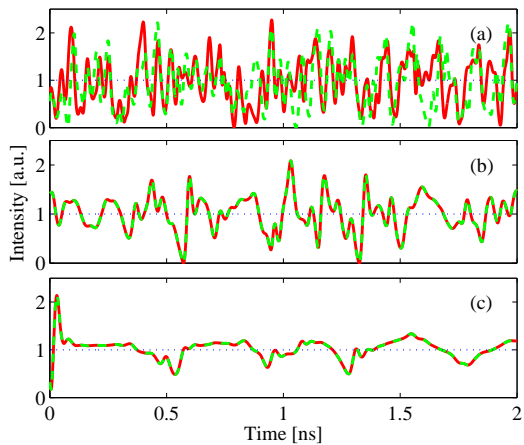}
\caption{Intensity $|E(t)|^2$ (dashed green curve) and $|E(t-T)|^2$ (solid red curve), normalized to unit average value, of the electric field at time $t$ and time $t-T$, together with the (constant) normalized intensity of the phase modulated driving field (dotted blue curve), for a) $T_m = 25$ ps, b) $T_m = 75$ ps, and c) $T_m = 150$ ps period of the random phase modulation.}
\label{Fig3}
\end{figure}

We notice now that the laser locks into a periodic trajectory even when the sampling rate is lower than the bandwidth of the chaotic input, approximately $15$ GHz in our numerical example. This is surprising because sampling the free-running chaotic waveform with a sampling rate significantly lower than its bandwidth is equivalent to selecting almost uncorrelated samples of the recorded chaotic trajectory. Stimulated by this observation, we decided to investigate a different configuration to achieve locking. A random phase modulation with uniform distribution from $0$ to $2 \pi$ is applied with $1/T_m$ rate to a continuous-wave master, which is then injected into the slave. The intensity of the master laser is not modulated. We set the forcing field $E_{\mathrm{ext}}(t)$ such that its intensity is equal to a fraction $\eta'$ of the average field intensity of the free-running slave. In Fig. \ref{Fig3} we show the results for $T_m = 25$ ps, $T_m = 75$ ps, $T_m = 150$ ps modulation time. For $T_m = 25$ ps the laser does not fully lock, although some periodic regularities show up. Full and robust locking is instead achieved with $T_m = 50$ ps and larger.  We have verified in this case as well that, when the laser locks, the periodic trajectory is still a global attractor of the system. Although the curves of Figs. \ref{Fig2} and \ref{Fig3} show qualitative differences, simple inspection on a scale shorter than the period would identify both as plots of genuine chaotic time series.

To obtain a quantitative characterization and a deeper understanding of the above observations, and to analyze the differences between sampled and random injection, we decided to geometrically characterize the attractor of the free-running laser by a numerical estimate of its Hausdorff dimension $D$ \cite{Ruelle}. Our estimate is obtained using an $N$-dimensional, with $N = 3 + 2 \Delta t/\delta t$, approximation of the infinite-dimensional phase space, consisting of the value of $N(t)$ and of samples of the real and imaginary parts of the electric field taken with $\delta t = 10$ ps sampling time in the interval $t - \Delta t \le t' \le t$. In this reduced phase space, we follow the evolution of the system in a long time interval, which we chose as the time $T$, the period of the controlled regime. During a simulation run of $T$ duration, we periodically register with period $t_s = 4 \, \delta t$ a full $N$-dimensional picture of the dynamical system, ending up with a set of points in the $N$-dimensional phase space (2500 points in our numerical example). We then numerically evaluate the correlation integral $C(r)$, defined as the average number of $N$-dimensional pairs with Euclidean distance less than $r$, averaged over all pairs, and we compute the exponent $\nu$ characterizing the small $r$ asymptotic power-law behavior of $C(r) \sim r^{\nu}$. For the free-running laser, the chaotic attractor has Hausdorff dimension $D \simeq \nu$ \cite{Procaccia}.

\begin{figure}
\includegraphics[width=8cm]{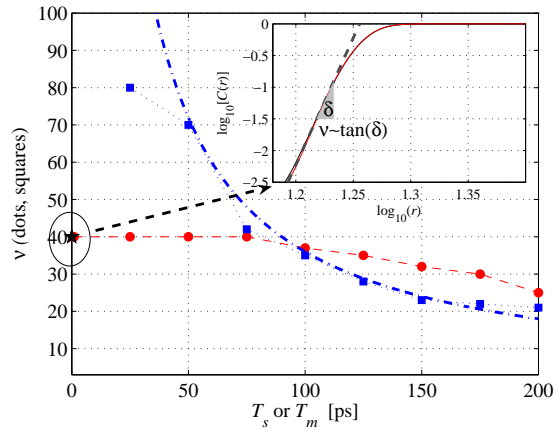}
\caption{Exponent $\nu$ for sampled injection (dots) vs. the sampling time $T_s$ ($T_s =1$ ps, ideal injection), and for random injection (squares) vs. the modulation period $T_m$. The value of $\nu$ of the free running laser is indicated with a star at $T_s =0$. Dot-dashed line: number of estimated degrees of freedom of the input drive vs. the modulation period $T_m$ for the case of random injection. In the inset, the curves of $\log_{10} \left[ C(r) \right]$ vs. $\log_{10} (r)$ for the free-running (solid black) and the locked case with ideal injection (solid red).}
\label{Fig5}
\end{figure}

For a laser locked to a stable periodic trajectory the attractor is rigorously one-dimensional, but the asymptotic exponent $\nu$ obtained using a time series of length $T$ has a meaning by itself. If, for a sufficiently long period $T$, the exponent $\nu$ is close to the value of the standalone laser, we may reasonably conclude that the periodic trajectory covers a large part of the chaotic attractor of the free-running laser, and it would be hardly distinguishable from a genuine chaotic trajectory on a time scale of the order of $T$ or shorter. In Fig. \ref{Fig5} we plot by red dots the values of $\nu$ vs. the sampling time $T_s$, showing with a black star at $T_s =0$ the exponent $\nu$ of the free-running laser and with a red dot at $T_s = 1$ ps that of the locked laser with ideal injection (i.e., sampling time equal to the numeric integration step, 1 ps). In the inset, we also plot the curves of $\log_{10}\left[C(r)\right]$ vs. $\log_{10} (r)$ for these two cases, showing no difference between the two. This confirms that it would be impossible, over a time scale of the order of $T$, to distinguish the periodic trajectory from the truly chaotic one. These results are not surprising. Indeed, the laser locks with high fidelity to a drive obtained from an arbitrary segment of the laser electric field periodically injected into it. Consequently, the ensemble of all possible periodic trajectories will be arbitrarily close, for a sufficiently long period $T$, to a coverage of the entire periodic attractor. It might therefore be safely stated that the manifold of the ensemble of all the possible periodic trajectories of the locked laser is the attractor of the free-running laser. This result holds whenever the control field is over-sampled, i.e. it is sampled with a sampling rate higher than its bandwidth, that is, in our case, for $T_s \le 75$ ps. In this regime, Fig \ref{Fig5} shows  that the exponent $\nu$ is always equal to the free-running value.

For $T_s > 75$ ps, the free-running laser trajectory used as a drive is under-sampled. This means that the sampled input has less information than the free-running chaotic trajectory. To give an example, going from $T_s = 75$ ps to $T_s = 150$ ps the number of independent samples is halved, hence the input would reduce by half the number of degrees of freedom. Figure \ref{Fig5} shows however that in this case the reduction by half of the number of degrees of freedom of the input produces a reduction of the exponent $\nu$ by only about 20 \%. This may be an indication of the property of the laser dynamics to recover part of the lost information of the drive, which is a natural trajectory of the system.

Figure \ref{Fig5} reports also the values of $\nu$ for a random-phase modulated forcing field, plotted vs. $T_m$ with blue squares. With random injection, the random drive does not belong by any means to the free running chaotic attractor. In our $N$-dimensional approximation of the phase space, the drive may be represented by points describing in their temporal evolution a curve belonging to a manifold of dimension $d \leq N$. Indeed, by the Nyquist-Shannon sampling theorem with time and frequency interchanged, a time-limited complex signal in an interval $\Delta t$, which is a point in our phase space, is fully characterized by its spectral components (modulus and phase) spaced by $\Delta B = 1/\Delta t$.  Although the number of spectral components to be included is virtually infinite, if the bandwidth of the drive is about $1/T_m$, the number of effective degrees of freedom of the drive is $d \simeq \Delta t/T_m$ (half, because of the constant intensity, of the degrees of freedom of an intensity and phase modulated signal). In the case of random phase modulation, the drive belongs to a manifold of dimensionality $d$ embedded into the phase space, which plays the role of the chaotic attractor of the sampled case. The ensemble of all possible laser outputs locked to the drive will belong to a manifold of dimensionality not larger than the dimension of the drive itself, because the laser, when locked, cannot create information. In Fig \ref{Fig5}, we compare the exponent $\nu$ estimated for the output with $d^* = 1.2 \, \Delta t/T_m$, represented with a dot-dashed line, where the (empirically chosen) 20 \% increment of $d$ accounts for the slightly larger bandwidth of the input if compared to a rigorously band-limited signal of bandwidth $1/T_m$. The comparison shows that the $d^* \simeq \nu$ hence that the laser output trajectory preserves all the degrees of freedom of the input. Notice that the value of $\nu$ corresponding to $T_m = 25$ ps is below the $1/T_m$ interpolation curve. This is because, in this case, the laser does not fully lock to the input, hence part of the degrees of freedom of the input are not reproduced by the output.

With random injection, the manifold of the random drive does not belong to the chaotic attractor of the free-running laser, hence the laser dynamics locked to the drive reproduce, at most, all the degrees of freedom of the drive itself. Consequently, the dependence of the exponent $\nu$ on the bandwidth, hence on the number of degrees of freedom, of the drive is stronger here than with sampled injection. To give an example, reducing the degrees of freedom of the drive by half reflects into a proportional decrease of the value of exponent $\nu$ of the output.

To conclude, we have shown that the injection of a periodic waveform can lock a chaotic system into a periodic trajectory belonging to its attractor.  The resulting process is periodic with a local chaotic signature, with a dimensionality that is different from the standalone process, depending on the specific injection scheme. Our analysis is specialized to a system obeying the modified L-K equations, appropriate, for example, to the description of an injected semiconductor laser driven to chaos by optical feedback from a short external cavity.  Here, the injection of a periodically replicated sampled copy of the laser output and a continuous-wave randomly-phase-modulated optical field lock the laser into a process that is locally equivalent to a chaotic dynamic. A careful analysis of the locked output reveals that with sampled injection the output waveform laser covers the same attractor of the standalone laser, whereas with random injection it belongs to a manifold of dimensionality equal to the number of degrees of freedom of the drive.

This work has been partially supported by the Italian Ministry of University and Research through the PRIN 2005 project No. 2005091255, \textit{Transmission system for optical chaotic cryptography}. Helpful discussions with Eugenio Del Re and Mauro Benedetti are acknowledged.


\begin{thebibliography}{99}

\bibitem{Ruelle} J.-P. Eckman and D. Ruelle, \rmp \textbf{55}, 617 (1985).

\bibitem{Ohtsubo_PRE} Y. Liu, N. Kikuchi and J. Ohtsubo, \pre \textbf{51} R2697--R2700 April 1995.

\bibitem{Watanabe} N. Watanabe and K. Karaki, \ol \textbf{20}, 1032--1034, May 1, 1995.

\bibitem{Gauthier} J. N. Blakely, L. Illing, and D. J. Gauthier, \prl \textbf{92}, 193901 (2004).

\bibitem{Ohtsubo} J. Ohtsubo, \textit{Semiconductor Lasers: Stability, Instability and Chaos}, Springer Series in Optical Sciences, Vol. 111, 2nd ed., 2008.

\bibitem{L-K} R. Lang and K. Kobayashi, \jqe \textbf{3}, 347 (1980).

\bibitem{Parameters} T. Heil, J. Mulet, I. Fischer, C. R. Mirasso, M. Peil, P. Colet, and W Els\"{a}{\ss}er, \jqe \textbf{38}, 1162 (2002).

\bibitem{Ippen} E. P. Ippen, D. J. Eilenberger, and R. W. Dixon, \apl \textbf{37}, 267 (1980).

\bibitem{Pecora} L.M. Pecora and T.L. Carroll, \prl \textbf{64}, 821 (1990).

\bibitem{Ott} E. Ott, C. Grebogi, and J. A. Yorke, \prl \textbf{64}, 1196 (1990).

\bibitem{Procaccia} P. Grassberger and I. Procaccia, \prl \textbf{50}, 346 (1983).



\end{thebibliography}
\end{document}